# Modeling Basal Ganglia for understanding Parkinsonian Reaching Movements

**K.N. Magdoom[1], D. Subramanian[1], V.S. Chakravarthy[1], B. Ravindran[2], Shun-ichi Amari[3]. N. Meenakshisundaram[1],**

[1]Department of Biotechnology, [2]Department of Computer Science and Engineering,

Indian Institute of Technology, Chennai, India.

[3]RIKEN Brain Science Institute, 1-2, Hirosawa, Wakoshi, Saitama, Japan.

## Abstract

We present a computational model that highlights the role of basal ganglia (BG) in generating simple reaching movements. The model is cast within the reinforcement learning (RL) framework with the correspondence between RL components and neuroanatomy as follows: dopamine signal of substantia nigra pars compacta as the Temporal Difference error, striatum as the substrate for the Critic, and the motor cortex as the Actor. A key feature of this neurobiological interpretation is our hypothesis that the indirect pathway is the Explorer. Chaotic activity, originating from the indirect pathway part of the model, drives the wandering, exploratory movements of the arm. Thus the direct pathway subserves *exploitation* while the indirect pathway subserves *exploration*. The motor cortex becomes more and more independent of the corrective influence of BG, as training progresses. Reaching trajectories show diminishing variability with training. Reaching movements associated with Parkinson's disease (PD) are simulated by (a) reducing dopamine and (b) degrading the complexity of indirect pathway dynamics by switching it from chaotic to periodic behavior. Under the simulated PD conditions, the arm exhibits PD motor symptoms like tremor, bradykinesia and undershoot. The model echoes the notion that PD is a dynamical disease.

## 1. Introduction

Reaching movements are to motor function, what the simple pendulum is to classical mechanics. Although straightforward to understand, reaching movements are of sufficient interest to researchers of motor function, since they are supported by a full array of motor areas in the brain. From a clinical point of view, reaching movements have diagnostic value since various motor disorders, like the Parkinson's disease (PD), manifest characteristic changes in reaching. PD patients are known to exhibit slower reaction times and movement times in simple movements aimed at targets (Brown and Jahanshahi, 1996). The first agonist burst in PD reaching movements is also observed to be weaker than in normals, resulting in longer and multi-staged reaching movements requiring multiple agonist-bursts. It has been suggested that part of the reason behind the longer movement times (MTs) in PD movements is that patients adopt a 'closed-loop' strategy to execute movements, which are executed in fast, open-loop fashion by normal subjects (Flowers, 1976). Though PD patients are capable of making fast ballistic movements, such performances resulted in impaired accuracy (Sheridan and Flowers, 1990). Another aspect of PD movements is the greater variability in movement end-point for larger movements. Thus bradykinesia, which refers to relative slowness of Parkinsonian movement, and 'closed-loop' mode of operation seems to be the strategy adopted by PD patients to make up for the inability to make consistent, large-amplitude movements.

In order to understand the changes in PD reaching movements just described, one has to understand the causal relationship between PD-related dopamine deficiency in basal ganglia (BG) and arm movements. Such understanding, if available, is best formulated in terms of a computational model. Bischoff (1998) presents a model of Parkinsonian arm control involving a simple reaching task and a reciprocal aiming task. Under PD conditions the model exhibits bradykinesia and also impaired ability to make sequential movements. Recently Cutsuridis and Perantonis (2006) modeled PD bradykinesia with an extensive model that includes, in addition to dopamine projections to BG and cortex but also, remarkably, dopamine projections to the spinal cord. The model successfully reproduces aspects of PD bradykinesia in terms of electromyographic (EMG) and movement parameters. Weaknesses of this model, however, include absence of explicit representations of BG nuclei and absence of a mechanism for learning reaching movements.

A perspective of BG function that has been gaining strength for over a decade now, is the idea that BG forms a neural substrate for reinforcement learning (RL), a branch of machine learning inspired by instrumental conditioning (Joel et al, 2002). Although a good number of BG models are not RL-based, most of them address only specific aspects of multitudinous functions of BG. Efforts are underway to explain the rich variety of BG functions solely within the RL framework (Chakravarthy et al, 2010).

A key feature of the proposed BG model consists in its interpretation of the role of the indirect pathway of BG, which has been given, in the past, a varied and tentative interpretation including: withholding of action (Albin et al, 1989; Frank, 2006), focusing and sequencing (Hikosaka et al, 2000), action selection (Redgrave et al, 1999), switching (Isoda & Hikosaka, 2008). We have been developing a line of modeling that hypothesizes that the indirect pathway subserves exploratory behavior (Sridharan et al, 2006). Thus the direct pathway and indirect pathway play complementary roles, whereby the direct pathway subserves exploitation while the indirect pathway supports exploration. Presence of complex dynamics in the indirect pathway justifies its putative role in exploration; and degradation of such complex activity to more regular forms of activity like synchronized bursts is hypothesized to contribute to impaired movement. Experimental evidence that is consistent with such a hypothesis is reviewed in Section 5. As a departure from the traditional description of BG functional anatomy according to which the direct pathway and indirect pathway support the 'Go' and 'NoGo' regimes respectively, we propose a third regime, *viz.*, the 'Explore' regime, which comes between the 'Go' and 'NoGo' regimes. This 'Explore' regime is also supported by the indirect pathway.

In this paper we describe a model of BG, which essentially belongs to the RL class of BG models. In this model, the dopamine (DA) signal is related to incremental changes in error between the target position and the position of the end-effector of the arm. DA level switches the transmission between the direct pathway and indirect pathway in BG (Clark et al, 2005). Output of BG is used as a corrective signal to the motor cortex (MC). This combined output of MC and BG is used to control the arm. As learning in MC progresses, motor cortex becomes gradually independent of BG, and begins to perform relatively independent of the modulatory influence of BG. Parkinsonian pathology is also captured naturally by the model through reduction of the dopamine level (Temporal Difference (TD) error, $\delta$) and by degrading the complex dynamics of indirect pathway.

The paper is organized as follows. Section 2 presents a brief background of BG structure and function. Section 3 describes the model architecture including training dynamics and measures of performance evaluation. Numerical simulations with the model training and testing under normal and Parkinsonian conditions are described in Section 4. A discussion of the work is given in the final section.

## 2. Background

### 2.1 Basal Ganglia Circuitry.

 The BG comprises of a group of subcortical nuclei, which form a highly interconnected network of modules. The caudate nucleus and the putamen, together referred as the striatum (STR), are the major input nuclei. Striatum receives inputs from a number of cortical areas and the thalamus. Another input port of BG, which is not however typically considered so, is the subthalamic nucleus (STN), which, like the striatum, also receives inputs from cortex. An internal module of BG, which is not directly connected to cortex or thalamus is the globus pallidus externa (GPe), which is thought to play a central role in BG according to more recent perspectives on BG circuitry (Obeso et al, 2006; Nambu, 2008). The BG has two output nuclei, globus pallidus interna (GPi) and substantia nigra pars reticulata (SNr). The output nuclei mainly target three nuclei: the thalamus, pedunculopontine nucleus and the superior colliculus. The activity of basal ganglia is modulated through constant feeds of the neurotransmitter dopamine (DA) from substantia nigra pars compacta (SNc) via the nigrostriatal pathway. The degeneration of neurons in SNc, whose axons form this pathway is known to cause idiopathic Parkinson's disease. Traditionally signal propagation through the BG is thought to occur via two alternative pathways viz., 1) the direct pathway which includes STR $\rightarrow$ GPi/SNr, and 2) the indirect pathway which consists of STR $\rightarrow$ GPe $\leftrightarrow$ STN $\rightarrow$ GPi. Dopaminergic transmission from SNc has a differential effect on striatal neurons according to striatum dopamine level: at smaller dopamine levels the indirect pathway is selected and an increase in striatal dopamine shifts the balance towards direct pathway, thereby increasing overall motor activity. Thus the indirect pathway is the normally active pathway. The balance is switched just before movement onset, when dopamine release to striatum activates the direct pathway (Clark et al, 2005).

**Functions of Basal Ganglia:**

Although for a long time, BG were thought to support motor functions exclusively, it is now recognized that BG have a role in cognitive, affective and autonomous function also. BG circuitry is involved in a great range of functions including: 1) reward based learning (Schultz 1998), 2) exploratory/navigational behavior (Packard & Knowlton 2002), 3) goal-oriented behavior (Cohen et al 2002), 4) motor preparation (Alexander 1987), 5) working memory (Cohen et al 2002), 6) timing (Buhusi and Meck, 2005), 7) action gating, 8) action selection (Redgrave, Prescott and Gurney, 1999), 9) fatigue (Chaudhuri and Behan, 2000) and 10) apathy (Levy and Dubois, 2005). In spite of the significant progress in our knowledge of BG at several levels, it is still not clear how such an overwhelming range of functions is supported by the same subcortical circuit.

**Reinforcement Learning – Dopamine:**

A key idea that opens doors to understanding of BG function is the idea that activity of dopaminergic cells in BG represents *reward signaling* (Schultz, 1998). More precisely, dopamine neurons are activated by rewarding events that are better than predicted, remain unaffected by events that are as good as predicted, and are depressed by events that are worse than predicted. Thus, dopamine signal seems to represent the *error* between predicted future reward and actual reward (Montague et al, 1996).

Interestingly, a quantity known as Temporal Difference error, analogous to this error between predicted and actual future rewards, plays a key role in reinforcement learning (RL) a branch of machine learning. This conceptual association enabled application of RL concepts (Sutton and Barto, 1998) to BG research (Joel et al, 2002). RL studies how an agent learns to respond to stimuli optimally without an explicit teacher; the agent's learning process is driven by reward/punishment signals that come from the environment in response to the agent's actions. Responses that result in rewards are reinforced and those that lead to punishment are avoided. Actor, Critic and Explorer are key components in a typical RL framework. The Critic is a module that estimates the reward-giving potential, the Value (V(t)), of the current state. The Actor uses the gradient in V(t) to choose actions that increase V(t). In the present model, when the gradient is absent or is too weak, choice of actions becomes increasingly stochastic. This stochasticity in choice of actions is identified with the Explorer.

### 3.0 The Complete Model: Arm, Basal Ganglia and Motor Cortex

Fig. 1 depicts the architecture of the arm-control system including the BG circuit, the motor cortex (MC), and the two-link arm model (AM). The motor task on which the system is trained consists of commanding the end-effector of the arm from the initial central position to one of the 4 surrounding targets. Information corresponding to the i'th target is coded in the Target Selection Vector (ξ) such that *i*'th component (ξ$_i$) is set to 1, while all the other components equal 0.   The Target Selection Vector is presented to both MC and BG (Fig. 1). Outputs of MC and BG, are combined to produce **g**, which represents the activations given to the 4 muscles of the two-link arm. The output of the BG may be regarded as a correction to the output of MC in controlling the arm. The basis of this correction is the error information associated with the relative position of the arm with respect to the target; this error is coded as the dopamine signal available to the BG. Thus in the model, the role of BG is two-fold: 1) to provide real-time corrective information to MC based on error information conveyed by nigrostriatal dopamine signal, and 2) to use this corrective signal to train the cortex on the motor task at hand.

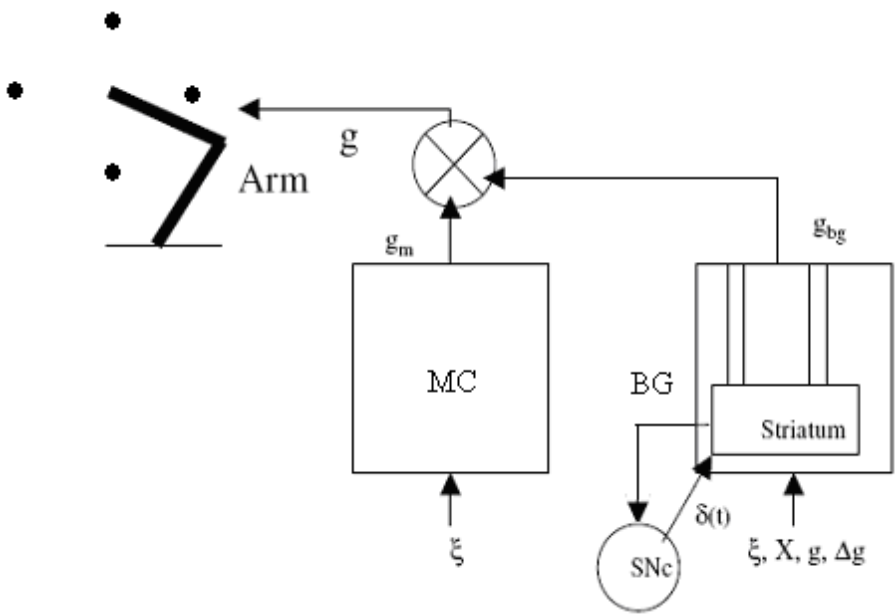

Figure 1: A schematic of the model components

The proposed model consists of three components: MC, BG, and the arm. The arm has to reach one of the 4 target locations. Each target is specified by a target selection vector, $\xi$, which is given as input to both the MC and the BG. In response to the input, the outputs produced by MC and BG are $g_m$ and $g_{bg}$ respectively. These outputs are linearly combined to produce $g$, which is given as activations to the muscles of the arm, calculated as:

$$g = \alpha g_m + \beta g_{bg} ,$$ (3.1)

where $\alpha$ and $\beta$ are coefficients that control the relative contributions of MC and BG to movement, as described below. A given value of $g$ puts the arm in a unique configuration. For a given value of $\xi$, BG output $g_{bg}$ is a highly labile quantity, which perturbs $g_m$ until the arm makes a successful reach. The value of $g_{bg}$ which results in a successful reach is used by MC for training itself. These processes are described below.

Inputs and Outputs of BG model:

Motor Cortex (MC): MC is modeled as a perceptron with $\xi$ as input and $g_m$ as output. Since the arm has four muscles (Fig. 1), $g_m$ is a 4-dimensional vector given as:

$$g_m = \tanh(W\xi + b)$$ (3.2)

where $W$ is the weight matrix, and $b$ is the vector of biases.

Basal Ganglia Model: The BG part of the model has 4 key components – 1) the Critic which is implemented in the striatum, 2) the direct pathway, 3) the indirect pathway and 4) the TD error, $\delta$, which represents the dopamine signal arising out of SNc.

<u>Critic</u>: For a given target position, $X_i^{tar}$, and the current position, $X$, of the end-effector of the arm ("hand"), the Critic computes the Value of the current position as:

$$V(X;X_i^{tar}) = A(1 - \frac{d^2}{R^2}) \qquad for \ \ d < R$$
$$\qquad\qquad = 0 \qquad\qquad otherwise$$

where $d = \left\| X - X_i^{tar} \right\|$                           (3.3)

where R represents the distance from the target over which the Value is non-zero.

<u>Dopamine Signal</u>: During the exploration of the arm in its planar workspace, if the arm accidentally strays close enough to the desired target, the BG receive a reward signal r(t). In line with standard RL literature, we define the TD error, δ, which represents the phasic activity of DA cells of SNc, as follows:

$$\delta(t) = r(t) + \gamma V(t) - V(t-1)$$                  (3.4)

where γ is the discount factor. The reward, r(t), is a sharp Gaussian of d, with mean 0, standard deviation, σ, and amplitude A, where d is the distance between the arm and the target. Thus reward is given when the arm comes very close to the target. δ is thought to be computed within the loop: striatum → SNc→ striatum (Fig. 1).

<u>Direct and Indirect Pathways:</u>

It is well known from functional anatomy of BG that striatal dopamine switches the transmission between the direct pathway and indirect pathway: the direct pathway is selected at higher values of dopamine, while the indirect pathway for lower values (Clark et al, 2005).

Selection of direct pathway is thought to facilitate movement ("Go") and selection of indirect pathway to withhold movement ("NoGo"). Between the 'high' and 'low' ranges of dopamine, which correspond to the classical "Go" and "NoGo" regimes, we posit an intermediate range, which corresponds to "Explore" regime. These three regimes operate in the current model as follows. In the "Go" case, the direct pathway is activated and $g_{bg}$ is updated such that the arm continues to move a little in the previous direction. In the "NoGo" case, the indirect pathway is activated and $g_{bg}$ is updated such that the arm shows a tendency to move a little in the direction opposite to the previous direction. In the "explore" case, again the indirect pathway is activated and $g_{bg}$ is updated in a random fashion unrelated to the previous increment in $g$. These mechanisms are embodied in the following equations:

$$if\,(\delta(t) > DA_{hi})$$
$$\qquad \Delta g_{bg}(t) = +\Delta g_{bg}(t-1) \qquad\qquad -"Go"$$
$$else\,if\,(\delta(t) > DA_{lo}\ \ and\ \ \delta(t) \leq DA_{hi})$$
$$\qquad \Delta g_{bg}(t) = \varphi \qquad\qquad\qquad -"E\exp lore" \qquad\qquad\qquad (3.5)$$
$$else\,//\,\,(\delta(t) \leq DA_{lo})$$
$$\qquad \Delta g_{bg}(t) = -\Delta g_{bg}(t-1) \qquad\qquad -"NoGo"$$

where φ is a random 4-dimensional vector calculated as in eqn. (3.12b). The threshold values $DA_{hi}$ and $DA_{lo}$ vary with training as,

$$DA_{hi} = a\beta; \quad DA_{lo} = -a\beta, \qquad\qquad\qquad (3.6)$$

where $a = 0.1$ and β is defined in eqn. (3.9) below,

and $g_{bg}$ is updated such that

$$g_{bg}(t+1) = g_{bg}(t) + \Delta g_{bg}(t)$$

$$(3.7)$$

Training MC:

The reaching movement driven by BG output as described by eqns. (3.1-3.7) proceeds until the end-effector comes within a radius of $R_{tol}$ from the target location. The value of 'g', which results in this successful reach, is used as target output of MC. Thus MC is trained by delta rule as follows:

$$\Delta W = \eta_m (g - g_m)\xi$$
$$\Delta b = \eta_m (g - g_m)$$

$$(3.8)$$

Two more quantities, $\alpha$ and $\beta$ (see eqn. 3.1), are updated as training proceeds. As MC becomes more and more skilled at the movement, the contribution of MC to movement increases, while that of BG decreases. This is achieved by making $\alpha$ a function of average reaching error, E, over all targets, when MC alone is used to drive the arm. Thus,

$$\alpha = \exp(-E); \quad \beta = 1 - \alpha.$$

$$(3.9)$$

Since E decreases with training, the contribution of MC, represented by $\alpha$, increases and the contribution of BG, denoted by $\beta$, decreases, with training.

Arm model:

Since BG dynamics is the focus of the paper, we chose an extremely simple model of arm dynamics. The arm consists of two joints with 4 muscles. The muscles are activated by $g$, a 4-dimensional vector: $g_1$ and $g_2$ activate the agonist and antagonist of the "shoulder" respectively, while $g_3$ and $g_4$ activate the agonist and antagonist of the "forearm" respectively. The "shoulder" and "forearm" joint angles, $\theta_1$ and $\theta_2$, respectively, are given by:

$$\theta_1 = \pi (g_1-g_2) \tag{3.10a}$$

$$\theta_2 = \pi (g_3-g_4) \tag{3.10b}$$

Thus in our simple arm model, the relationship between muscle activations and arm configuration is a static one.

<u>Modeling Parkinsonian Dynamics:</u>

Two mechanisms are used in the present model to simulate Parkinsonian conditions:

1. Reduced dopamine: Reduced dopamine conditions are simulated by putting a ceiling on the DA level, i.e., on $\delta$. This ceiling is imposed on $\delta(t)$, which is governed by eqn. (3.4), as follows:

$$\delta_{new} = \min(\delta, DA_{ceil}) \tag{3.11}$$

where min(x,a) is defined as:

$$y = x, \quad for \ \ x < a$$
$$= a, \quad for \ \ a \le x$$

Parkinsonian disease progression is simulated by gradually decreasing $DA_{ceil}$. For every value of $DA_{ceil}$, MC is trained for several epochs before reducing $DA_{ceil}$ further. In the simulations of Parkinsonian conditions described in the following section, we start with $DA_{ceil} = 0.5$ and reduce it in steps to -0.5.

2. Reduced complexity in the Explore regime: In order to vary the complexity of indirect pathway dynamics during the Explore regime, we use a chaotic map to calculate the random vector $\varphi(t)$ used in eqn. (3.5). Each component of $\varphi(t)$ is calculated using a logistic map as follows:

$$x_i(t) = K \, x_i(t-1)(1 - x_i(t-1)) \qquad\qquad\qquad (3.12a)$$

$$\varphi_i(t) = B \, x_i(t) \qquad\qquad\qquad\qquad\qquad (3.12b)$$

where $B$ denotes a scaling parameter.

The map parameter K controls the transitions between fixed-point ($0 \le K < 3$), periodic (for ($3 \le K < 3.57$ (approx)), and chaotic ($3.57 \le K \le 4$) behaviors of the logistic map (May, 1976). However, note that even in the "chaotic regime" ($3.57 \le K \le 4$) there are the so-called islands of stability, small ranges of K where the map exhibits periodic behavior. However, these islands are not likely to be detected in our simulations since we have not scanned the space of K at sufficiently high resolution. In the simulations of the following section, we designate K=4 to correspond to normal condition, and use smaller values of $K$ down to 3, to simulate PD-related degeneration.

## 4.0 Simulations: Normal and PD reaching movements

4.1. <u>Normal reaching movements</u>: The model described in the previous section is used to reach the 4 targets shown in Fig. 1. Training simulations are run for 20 epochs, where each epoch consists of reaching (or making time-limited attempts to reach) all the 4 targets. The weights of MC are randomly initialized between -0.5 and 0.5. Each reaching movement lasts for at most 100 time-steps, or until the arm serendipitously comes sufficiently close to the target. Thus reaching movements are made once towards each of the 4 targets in one epoch. This process is repeated for 20 epochs, at the end of which MC is almost completely trained. Even if training is continued beyond 20 epochs MC error does not reach 0, but fluctuates around a small positive value. The labile influences coming from BG to train MC play a dual, and mutually conflicting, role. This variability, on one hand, is necessary to explore the output space and discover rewarding increments to muscle activations. The same variability, on the other hand, prevents the MC from learning further once a low error value is reached. It is for this reason that we increase α (MC's contribution to movement), and decrease β (BG's contribution), as a function of training error, as training progresses. Numerical values of various parameters used in the model are listed in Table 1.

Table .1: Numerical values of parameters used in the model equations of Section 3.

| Parameter | Value | Description |
|-----------|-------|-------------|
| A | 2 | Amplitude of the reward and value functions |
| R | 3 | Spread of the value function |
| $\gamma$ | 1 | Discount factor |
| $R_{tol}$ | 0.3 | Radius of the tolerance circle |
| a | 0.1 | Scaling factor for $DA_{hi}$ and $DA_{lo}$ in BG function |
| $\eta_m$ | 0.2 | Learning rate of MC |
| $\sigma$ | 0.03 | Standard deviation of the Gaussian used in reward calculation |
| B | 0.04 | Scaling factor for logistic map function |

The parameters of Table 1 are chosen by experimentation keeping in view various trade-offs involved. For example, *a* controls the DA thresholds in eqn. (3.6). Larger values of *a* increases time spent in exploration. Similarly B controls amplitude of exploration. A larger value of $R_{tol}$ increases probability of a successful reach, but worsens reaching error. Variation of these parameters within a small range around the currently used values did not exhibit any sudden unexpected changes in system behavior. However, a systematic sensitivity analysis of the above parameters could form part of a separate study.

Evolution of reaching performance over the epochs is characterized by three metrics: 1) performance error of MC, 2) reaching duration, and 3) path variability. Fig. 2 shows the trajectories of the arm in the first epoch. Since the MC is yet untrained the arm makes long, wandering movements to reach the target. Arm trajectories are nearly straight in the last (20th) epoch (Fig. 3). Fig. 4 shows the reaching movements made by the arm under the sole influence of MC, without the BG contribution. This is done by setting $\alpha=1$ and $\beta = 0$. Note that the perturbative influence of BG is absent in this case. Note that though average reaching duration appears to decrease in the mean with learning, the trend does not seem to be significant when the error bars are considered (Fig. 5). Note that the error bars in all figures denote standard deviation. As the MC learns to reach, the initial movement, which is driven by the MC, arrives closer and closer to the target; thus the time-consuming wandering search for the target is reduced as learning progresses. For the same reason, path variability is also reduced as training progresses (Fig. 6). Naturally, since the goal of training is to train the MC to reach, MC reaching error decreases with epochs (Fig. 7).

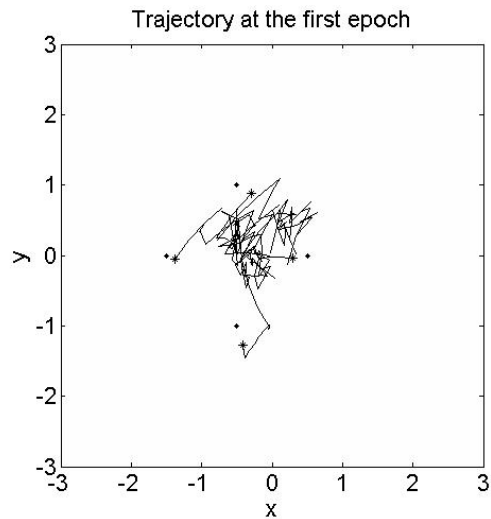

Figure 2: Trajectories of reaching movements during the first epoch.

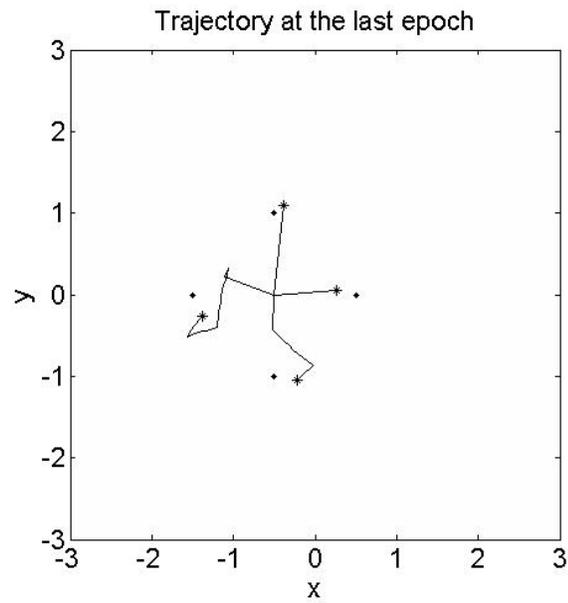

Figure 3: Trajectories of Reaching movements made during the last (20<sup>th</sup>) epoch.

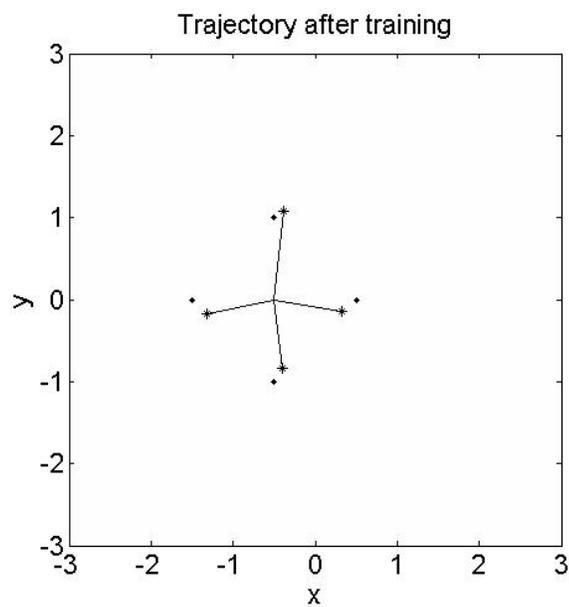

Figure 4: Reaching movements made to the four targets after training (MC alone, no BG contribution)

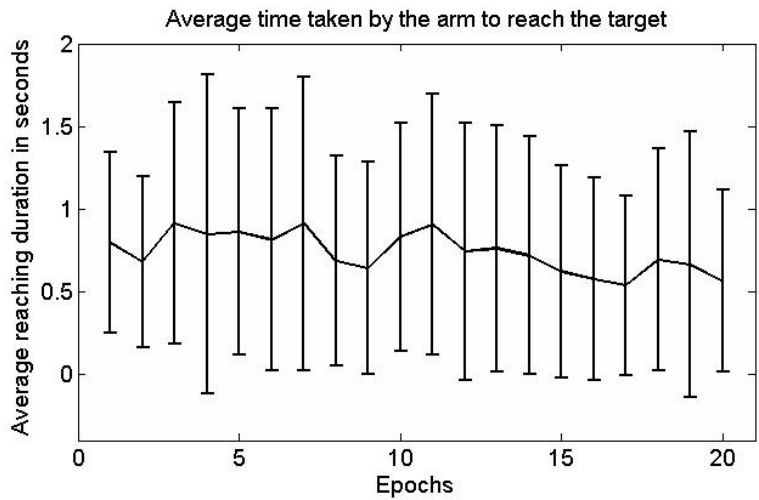

Figure 5: Variation of Average Reaching Duration with learning in normal conditions.

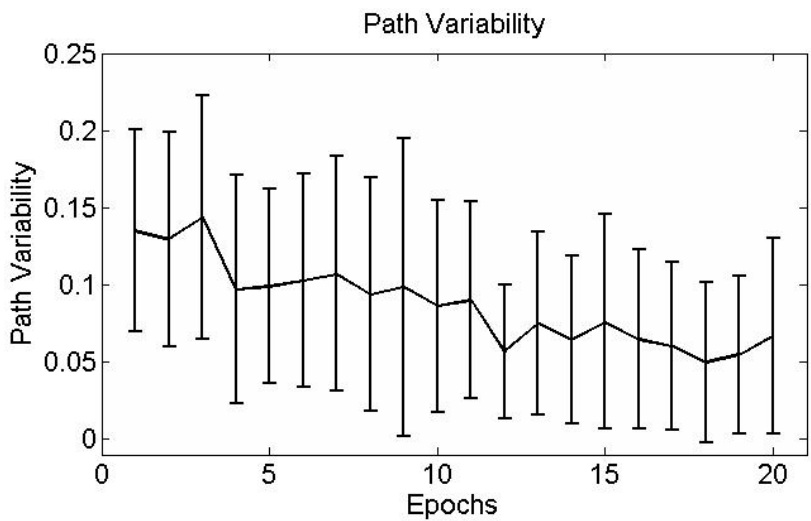

Figure 6: Variation of Path Variability with learning in normal conditions.

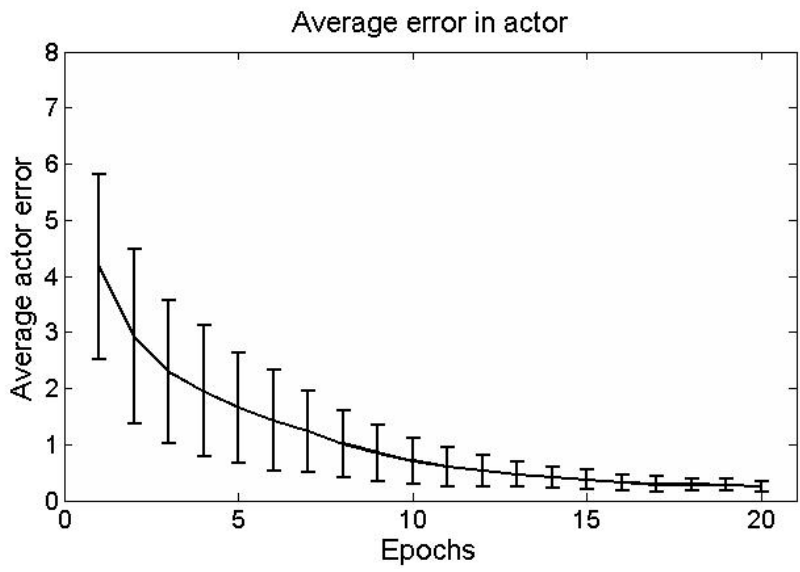

Figure 7: Variation of output error of MC with learning in normal conditions.

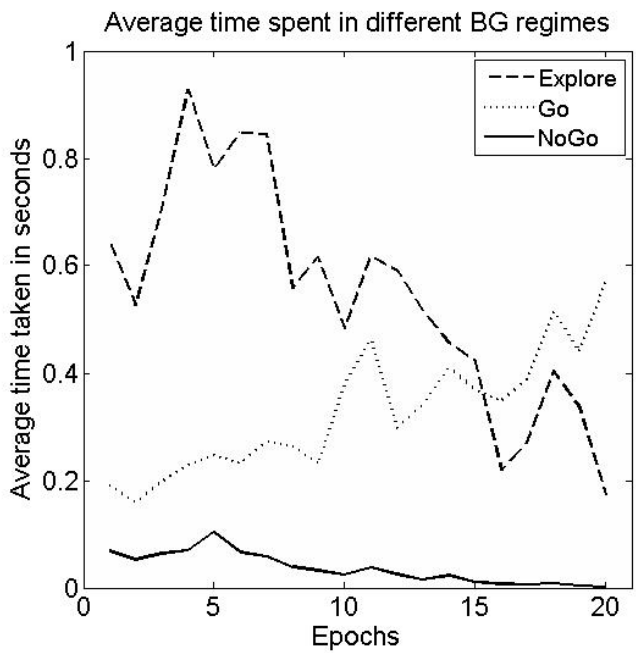

Figure 8: Average time spent in various BG regimes

The classical description of the function of direct pathway and indirect pathway associates direct pathway with movement facilitation, and indirect pathway with movement inhibition. In the present model, we propose that the dynamics of the 'Go' and 'NoGo' regime are opposite to each other: the respective changes in BG output in the two regimes have opposite signs. However, it may be argued that a simpler way to implement NoGo regime is to let the BG output remain unaltered. We implemented this variation of NoGo regime also and found that the results were qualitatively same (see Appendix-II). Therefore, we continue with the formulation of regimes as depicted in eqns. (3.5) and consider their consequences in Parkinsonian conditions in the next section.

## 4.2. PD reaching movements:

Simulations of PD-related pathology are based on three different types of models. In PD model – Type A, both dopamine reduction (-0.5 < $DA_{ceil}$ < 0.5) and reduced complexity of indirect pathway dynamics (3 < K < 4) are incorporated. In PD model - Type B, only dopamine reduction is implemented (i.e., -0.5 < $DA_{ceil}$ < 0.5, and K = 4). In PD model - Type C, only reduced complexity of indirect pathway dynamics (3 < K < 4) is incorporated with no reduction in dopamine ($DA_{ceil}$ = 0.5).

We define a few metrics to characterize reaching performance in PD conditions also. These are: 1) Undershoot factor, which quantifies the extent by which the final position of the arm undershoots the target, 2) tremor factor, which quantifies the "tremor" seen in the arm movements, and 3) average velocity to quantify bradykinesia. Formal definitions of these metrics are given in the Appendix-I.

Since loss of dopaminergic cells in SNc is the etiology of idiopathic PD, it would be natural to describe the degree of PD by percentage loss of DA cells. In this simulation, the degree of PD pathology is expressed by the quantity $D_{ceil}$, which clamps the DA signal $\delta$. We now define a quantity, $P_{DA}$, which represents percentage of DA cell loss, and relate it to $D_{ceil}$. Note that $\delta$ can take both positive and negative values, typically varying between -0.5 and 0.5 in the simulations. When $P_{DA}$ = 0, $D_{ceil}$ can take its highest value of 0.5, and when $P_{DA}$ = 1, $D_{ceil}$ takes its lowest value of -0.5. Thus we have, $D_{ceil} = 0.5 - P_{DA}$. For a given trial, there are a total of 20 epochs for each 5% of DA loss and each epoch lasts for a maximum of 100 iterations, if the arm freezes for more than 10 time steps then the reach is terminated. There are a total of 10 trials for each DA level. Trials represent repeated simulation for the same DA level. Such repetition is necessary to examine the level of variability in reaching.

In the three types of PD simulations, we start with the MC fully trained under normal conditions (as in Section 4.1), and continue to train it under the pathological conditions by varying $P_{DA}$ and

K. Variation of various metrics like undershoot etc. presents three possible scenarios of PD disease progression.

PD model – Type A:

As mentioned above, in this PD model, both dopamine reduction (-0.5 < DA$_{ceil}$ < 0.5) and reduced complexity of indirect pathway dynamics (3 < K < 4) is incorporated. DA$_{ceil}$ and K are related to P$_{DA}$ as follows: $D_{ceil} = 0.5 - P_{DA}$ and $K = 4 - P_{DA}$.

PD patients are known to often undershoot target in reaching performance (Van Gemmert et al, 2003). This is clearly seen in fig. 9, where undershoot worsens with increasing loss of DA cells (P$_{DA}$). At around 50% loss of DA cells, undershoot nearly reaches its minimum and does not change significantly henceforth. Fig. 10 shows a snapshot of reaching trajectories with undershoot. Note that apart from undershooting the target, there is also a large error in reaching direction.

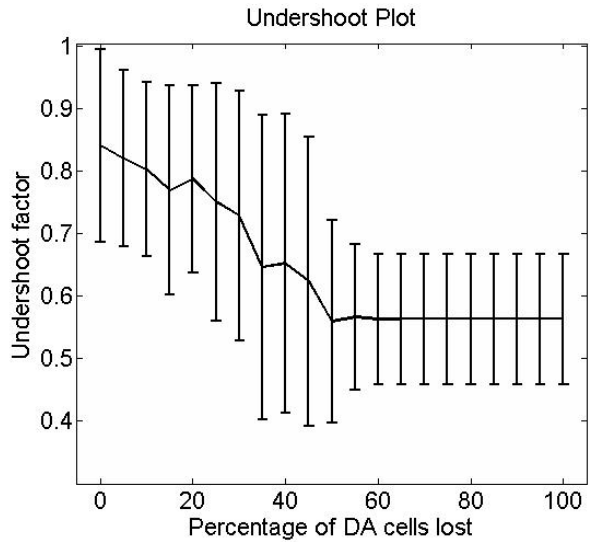

Figure 9: Variation of Undershoot factor with P$_{DA}$, for Type-A PD model

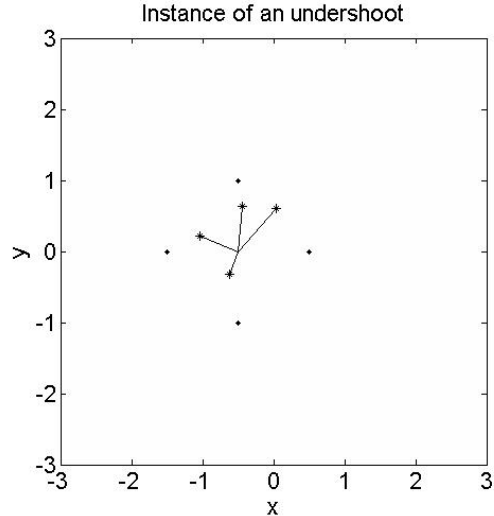

Figure 10: Instance of Undershoot in case of Type-A PD model ($P_{DA}$ = 100%)

Tremor also increases with increasing $P_{DA}$ up to about $P_{DA}$ =50%; henceforth it quickly drops to 0 at $P_{DA}$ =60%, and remains at 0 for larger values of $P_{DA}$ (Fig. 11). This development may be accounted for as follows. As $P_{DA}$ is increased, $DA_{ceil}$ is also reduced, and $\delta$ spends more and more time in "Explore" and "NoGo" regimes. Such exaggerated exploration, occurring in place of a straight target pursuit corresponding to "Go" regime, seems to manifest as "tremor". As $P_{DA}$ is increased further, $\delta$ is always confined to "NoGo" regime, irrespective of the actual performance of the arm. Thus the arm enters a relatively "frozen" state with no tremor. Ramifications of this change can be seen in average velocity also. Average velocity decreases with increasing $P_{DA}$, reaching small average velocity at about $P_{DA}$ =50, and remaining there for larger values of $P_{DA}$ (Fig. 12). Although the disease pathology is confined to BG, performance error of MC gradually increases with increasing $P_{DA}$ and saturates at about $P_{DA}$ =50 (Fig. 13). Thus undershoot and average velocity seem to show a common pattern: a nearly gradual worsening up to about $P_{DA} = 50$, and a subsequent relatively frozen condition marked by a paucity of movement. However, tremor gradually increases to a peak value, before falling rapidly around $P_{DA} = 50$. These patterns seem to be reflected in the variation of times spent in various regimes (Fig. 14). Variation of time spent in the Explore regime seems to resemble variation of tremor, whicle undershoot and average velocity variation seems to follow the variation of time spent in Go regime (Fig. 14).

Normal reaching is accompanied by a large, initial agonist burst, followed by an antagonist burst, which sometimes, is followed by a second, smaller agonist burst. (Fig. 15 top). Note that all agonist burst plots correspond to the variation of the first component of the muscle activation vector (representing the shoulder) as the arm reaches target 1. We have not included other components, since they also show a similar behavior. Time zero in all agonist burst plots corresponds to the start of the reaching movement. Such biphasic response seen in normal case, does not appear in PD –Type A results which show a nearly monotonic build up of activity (Fig. 15 bottom).

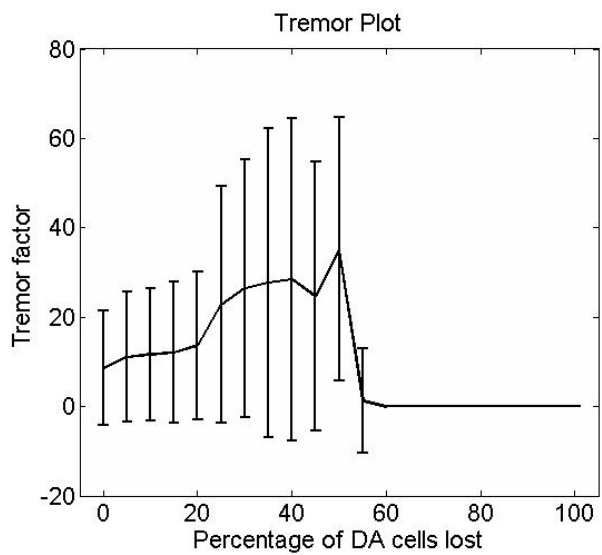

Figure 11: Tremor factor in case of Type-A PD model

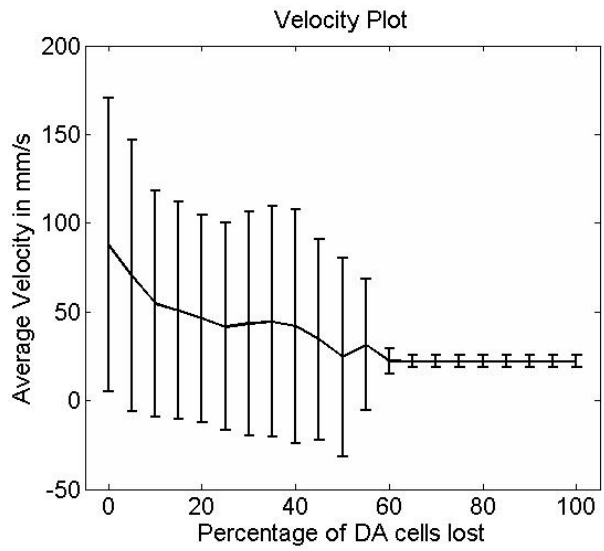

Figure 12: Average velocity of the arm in case of Type-A PD model

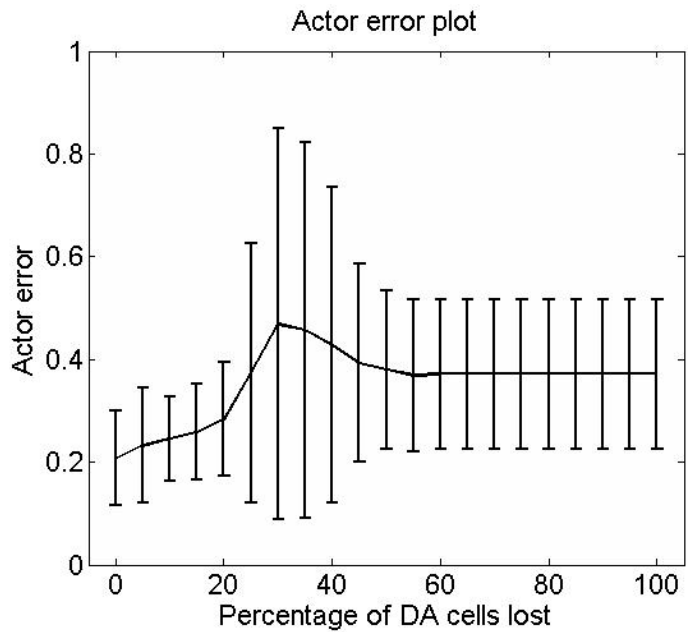

Figure 13: Actor or MC error in case of Type-A PD model

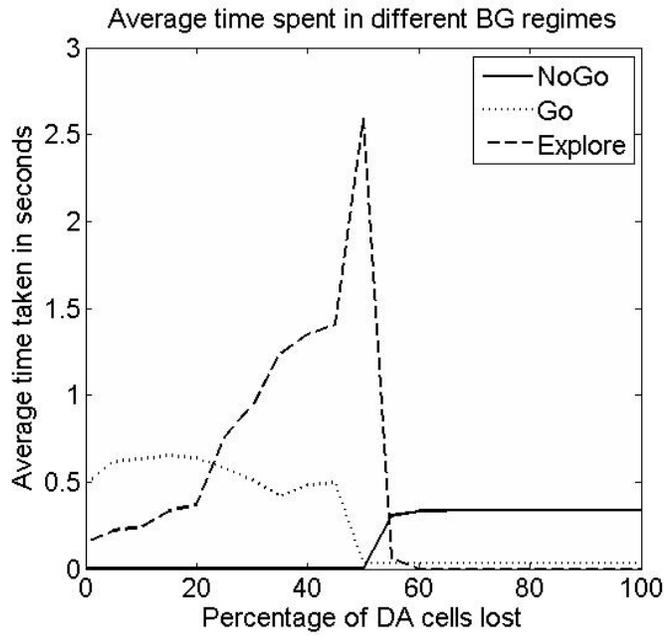

Figure 14: Average time spent by BG in various regimes for Type-A PD model

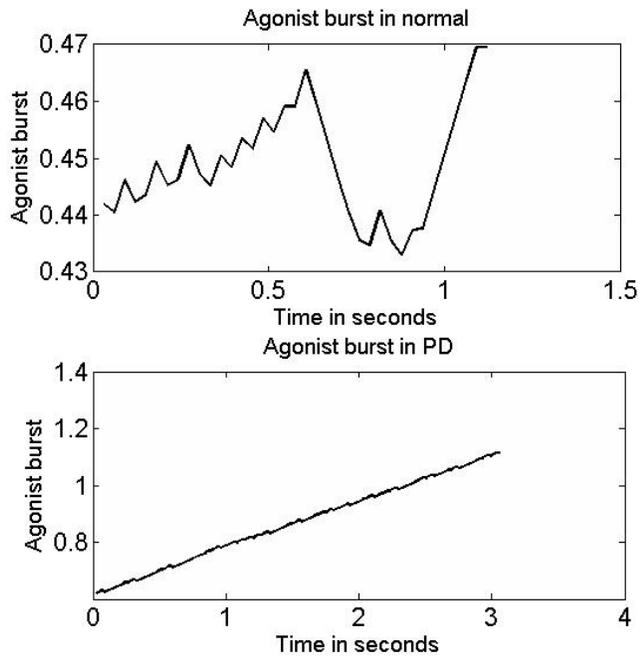

Figure 15: Agonist burst. Normal (top) and PD model-Type A (bottom). Normal plot corresponds to target number 1, and epoch number 16. PD plot corresponds to DA loss of 30%, and target number 1.

PD model – Type B:

As mentioned earlier, in this PD model, only dopamine reduction ($-0.5 < DA_{ceil} < 0.5$; K $= 4$) is incorporated. $DA_{ceil}$ is related to $P_{DA}$ as follows: $D_{ceil} = 0.5 - P_{DA}$ .

Although the general trends are similar to those seen in case of PD model-Type A, there is an important difference. For example, if we consider the variation of undershoot with $P_{DA}$, in case of Type A model, there is a gradual reduction followed by saturation. However, in case of Type B model, undershoot remained nearly constant and fell drastically at $P_{DA}$ value of about 50%, without much subsequent variation (Fig. 16). Fig. 17 shows a snapshot of undershoot in this case. Tremor also remains nearly constant until $P_{DA}$ =50%, falling abruptly to 0 thereafter (Fig. 18). However, tremor exhibits a sharp transient rise at $P_{DA}$ =50% before it falls to 0. A similar step-like change is observed in average velocity also (Fig. 19). However, actor (MC) error shows insignificant variation with $P_{DA}$ (Fig. 20). In this case too, variation of symptoms reflect variation of time spent in various regimes. For instance, as in the previous case, variation of average velocity and undershoot resemble variation of time spent in Go regime, while variation of tremor resembles Explore regime (Fig. 21). In this case too, agonist burst shows a monotonic variation (Fig. 22, bottom), compared to a biphasic response of normal case (Fig. 22, top).

Thus, unlike the Type A model, in Type-B model, symptoms show a step-like variation, with the symptom remaining constant up to a critical value of $P_{DA}$ =50%, thereafter transitioning to a permanently worse state. This can perhaps be accounted as follows: the loss of DA neurons might be compensated by the intact indirect pathway dynamics and this balance is perhaps disturbed when $P_{DA}$ reaches a critical level (in this case 50%).

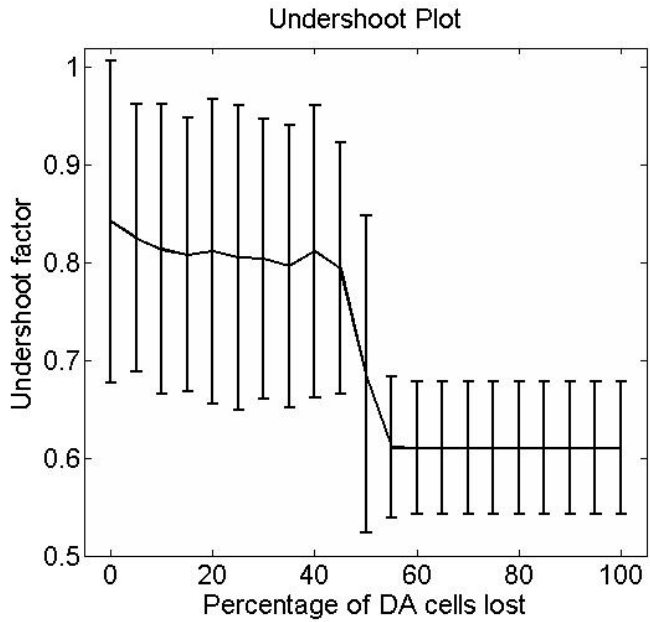

Figure 16: Variation of Undershoot Factor with percentage DA cell loss in case of Type-B PD model

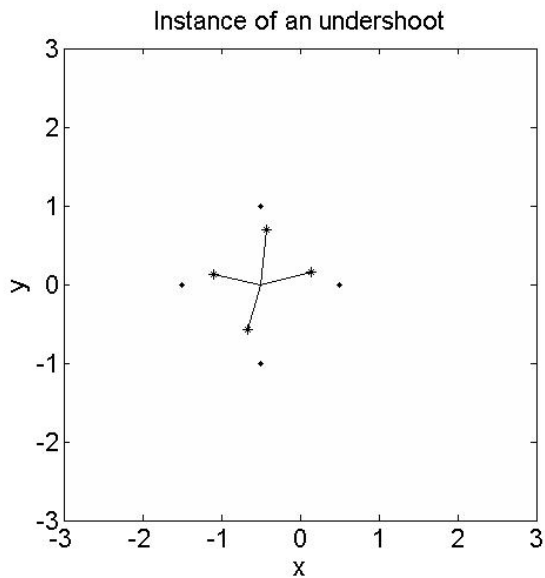

Figure 17: A snapshot of undershot reaching movement in case of Type-B PD model ($P_{DA}$ = 100%)

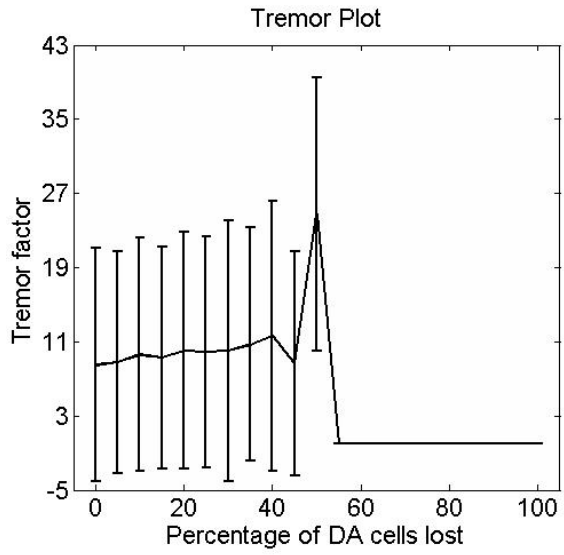

Figure 18: Variation of Tremor with percentage DA cell loss in case of Type-B PD model

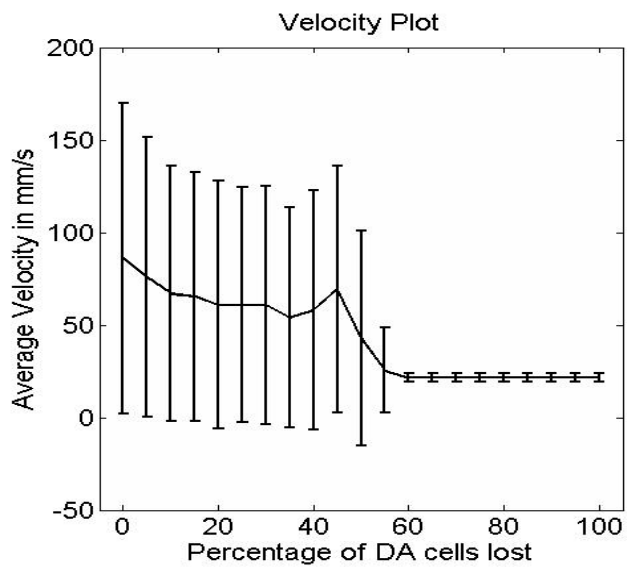

Figure 19: Variation of Average Velocity with percentage DA cell loss in case of Type-B PD model

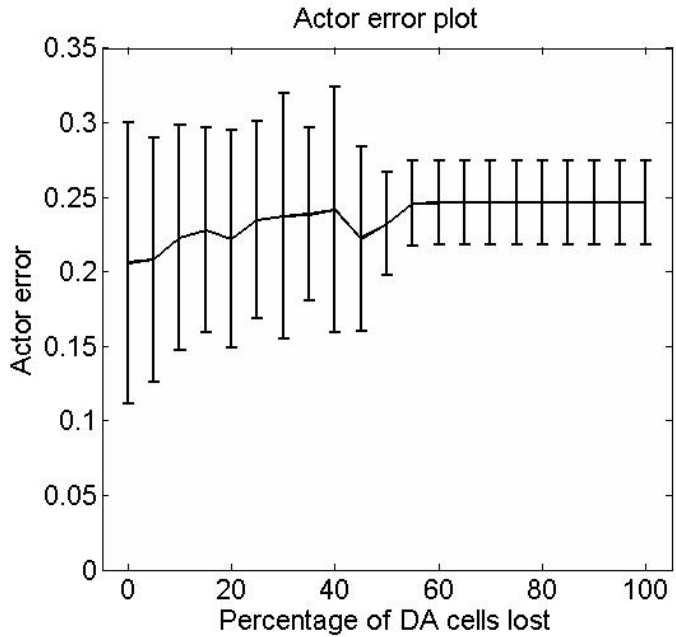

Figure 20: Variation of Actor or MC error with percentage DA cell loss in case of Type-B PD model

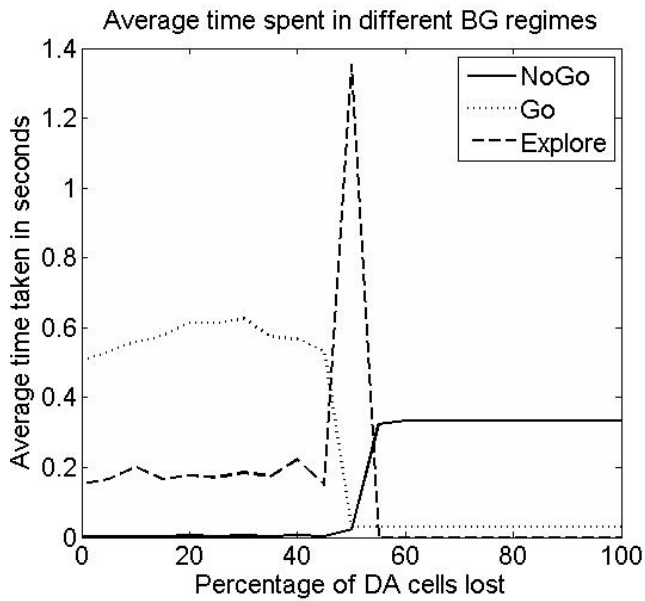

Figure 21: Average time spent by BG in various regimes for Type-B PD model

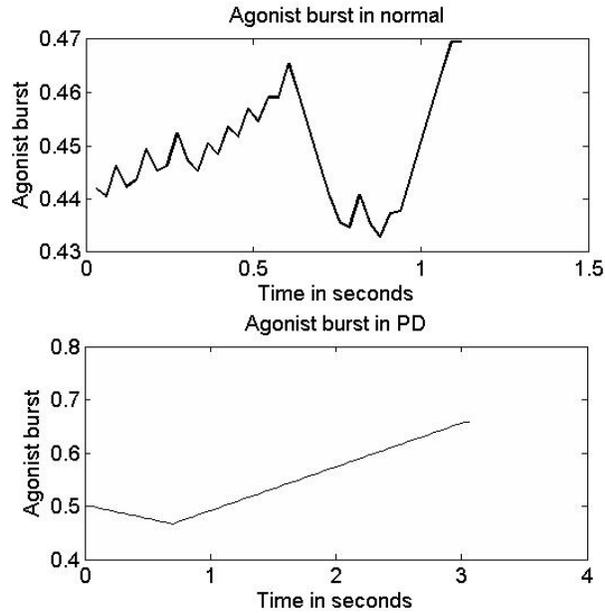

Figure 22: Agonist burst. Normal (top) and PD model-Type B (bottom). Normal plot corresponds to target number 1, and epoch number 16. PD plot corresponds to DA loss of 30%, and target number 1.

PD model – Type C:

As mentioned in this PD model, only reduced complexity of indirect pathway dynamics ($3 < K < 4$) is incorporated with no reduction in dopamine ($DA_{ceil} = 0.5$). $DA_{ceil}$ is fixed at 0.5, and K is related to $P_{DA}$ as follows: $K = 4 - P_{DA}$.

In Type A simulations, we have seen a gradual variation of symptom, followed by saturation at $P_{DA}$ of about 50%. In Type B, we have seen a nearly constant profile up to $P_{DA}$ of about 50% followed by a sudden shift to another plateau. In Type C, we see a generally gradual variation of symptoms (Undershoot factor, Fig. 23; Tremor factor, Fig. 25; Average Velocity, Fig. 26) with no sharp transition at $P_{DA}$ of about 50%. Since $DA_{ceil}$ is fixed, $\delta$ is allowed a full, unconstrained variation. Thus the sharp transitions between the three regimes do not occur here. Whatever impairment of symptoms observed, is due to degradation of complexity of exploration (reduction in K). Note that the boundaries between regimes are not fixed, but vary as a function of Actor error (see eqn. (3.6)). The span of Explore regime is wider for higher Actor error. Thus, Actor error increases as $P_{DA}$ increases (Fig. 27). As a consequence, for larger values of $P_{DA}$, the system spends more time in Explore regime than Go regime (Fig. 28). Since $\delta$ never or rarely drops too low, NoGo regime is hardly selected (Fig. 28). In this case too, agonist burst shows a monotonic variation (Fig. 29, bottom), compared to a biphasic response of normal case (Fig. 29, top).

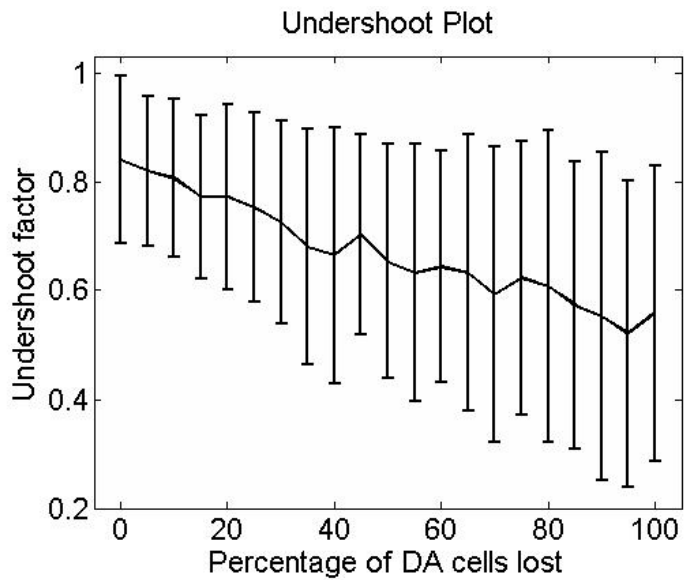

Figure 23: Variation of Undershoot Factor with percentage DA cell loss in case of Type-C PD model

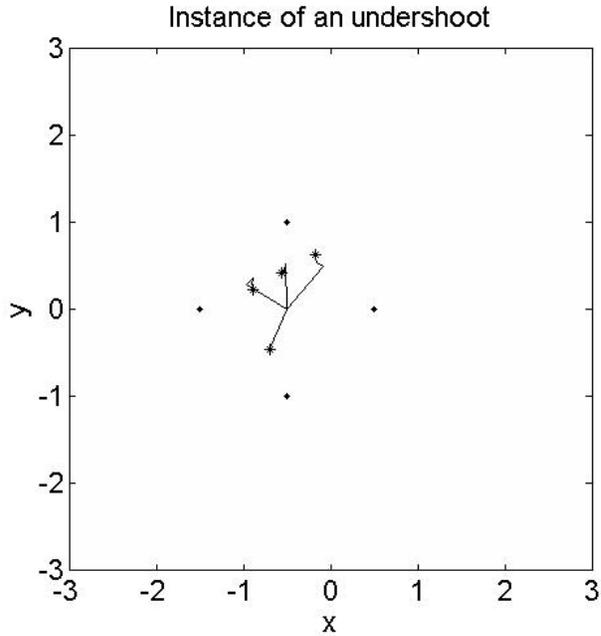

Figure 24: A snapshot of undershot reaching movement in case of Type-C PD model ($P_{DA}$ = 100%)

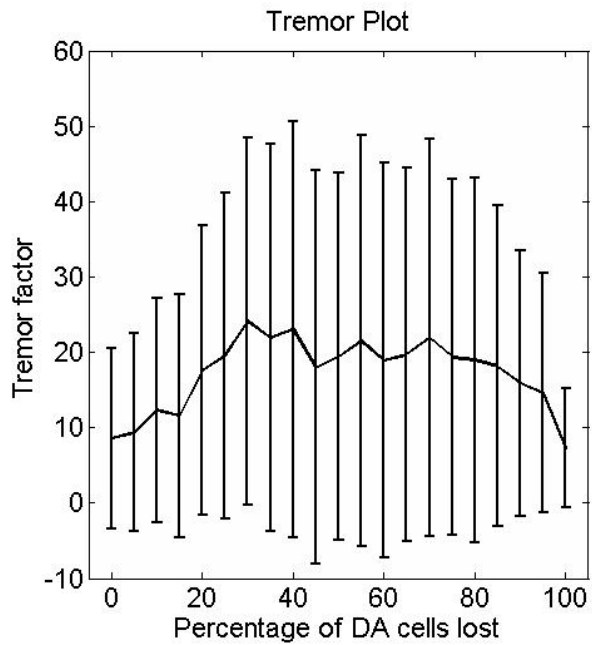

Figure 25: Variation of Tremor with percentage DA cell loss in case of Type-C PD model

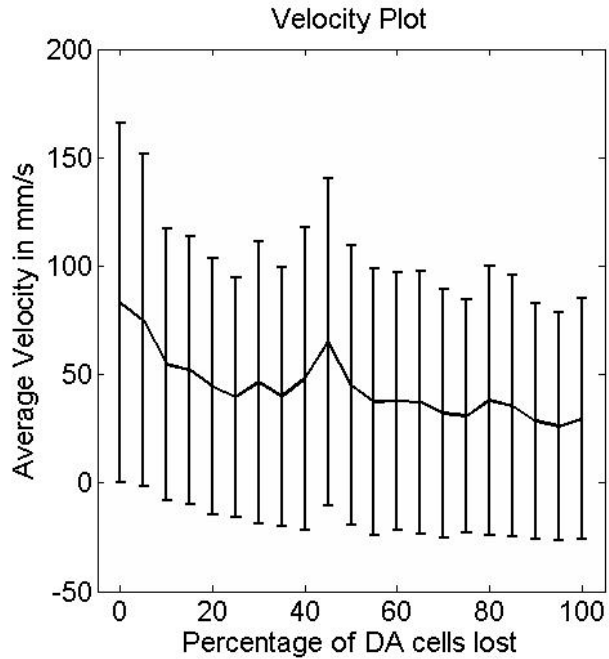

Figure 26: Variation of Average Velocity with percentage DA cell loss in case of Type-C PD model

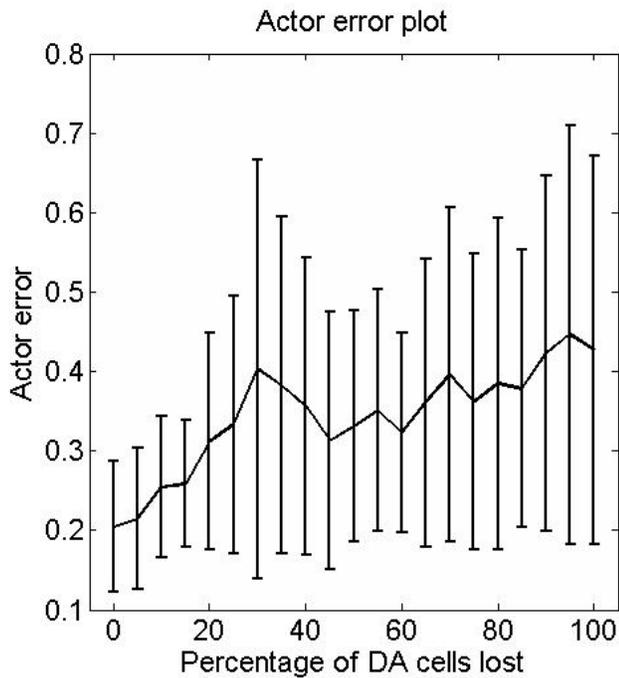

Figure 27: Variation of Actor or MC error with percentage DA cell loss in case of Type-C PD model

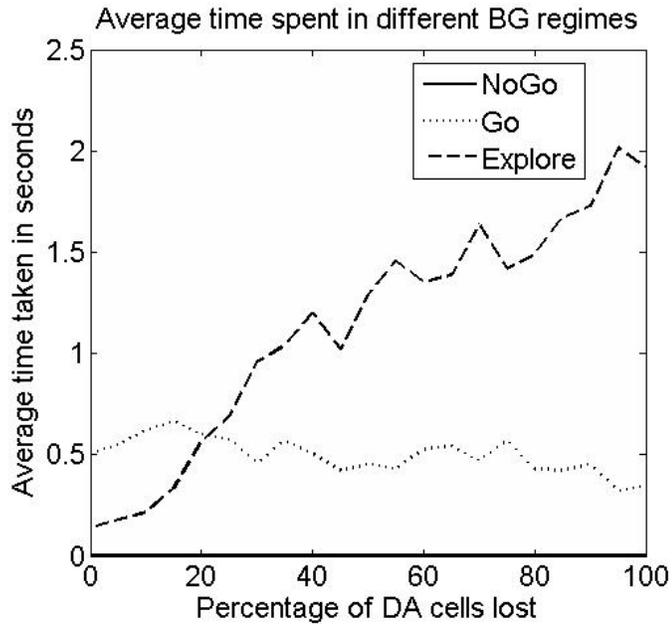

Figure 28: Average time spent by BG in various regimes for Type-C PD model

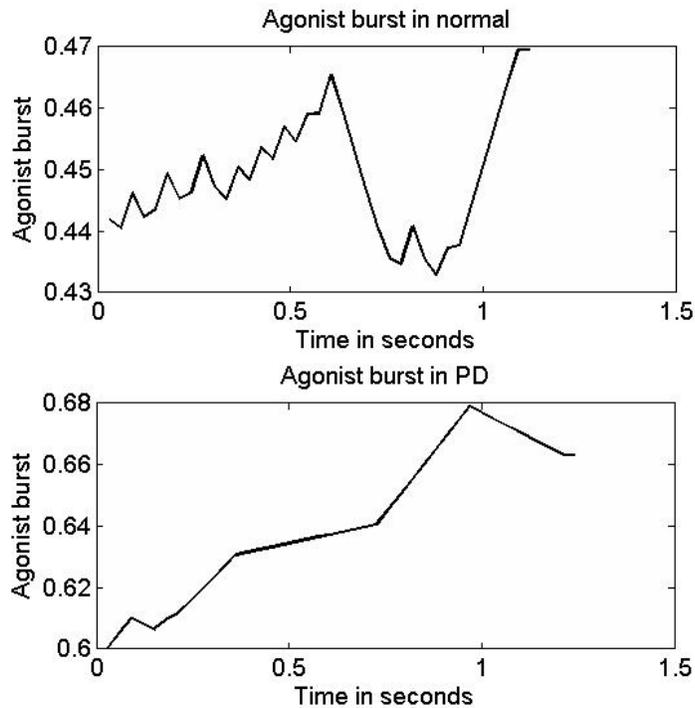

Figure 29: Agonist burst. Normal (top) and PD model-Type C (bottom). Normal plot corresponds to target number 1, and epoch number 16. PD plot corresponds to DA loss of 30%, and target number 1.

5.0 **Discussion:**

We present a model of Parkinsonian reaching dynamics. The model consists of MC, BG and a two-link arm. The BG model is cast essentially in the framework of reinforcement learning, though we depart radically from the interpretation of neural substrates of various RL components. In line with actor-critic type of BG models, we interpret the temporal difference error as the DA signal. The value function, which is thought to be computed in the striatum, is not learnt but predefined in terms of distance of the arm's end-effector and the target. DA signal switches transmission between direct pathway and indirect pathway. Thus BG output is dominated by direct pathway or indirect pathway activity, depending on the magnitude of the DA signal. BG output in combination with MC output controls the arm. Thus the perturbative corrections from BG, and the DA signal, together help MC to learn to reach. MC's dependence on BG gradually diminishes as training progresses. Thus, in the present model, BG discover the correct output by reward-related dynamics, and transfers the knowledge to MC. A similar scenario of sequential learning between BG and cortex was described in experimental literature. Studies on different time-courses of learning in basal ganglia and prefrontal areas exhibit a similar sequencing (first basal ganglia and then prefrontal) in saccade related behavior in monkeys (Pasupathy & Miller 2005).

The DA signal in the present paper does not distinguish between the two forms of DA release from mesencephalic dopamine centers reported in experimental literature: the phasic release which acts on a time-scale of seconds, and tonic release, which acts over a few minutes (Dreher and Burnod, 2002). Phasic release is linked to the difference in expected future reward and actual reward, a quantity described in RL literature as the Temporal Difference error. Both tonic and phasic dopamine releases are thought to have differential roles in efficient updating of working memory information in the prefrontal cortex. Tonic DA is thought to increase stability of maintained information in PFC by increasing the signal-to-noise ratio of the pattern with respect to background noise. By contrast, phasic DA is thought to control when an activity has to be maintained or when it must be updated (Cohen et al, 2002). The possibility of interaction between tonic and phasic dopamine has also been considered. It has been suggested that tonic dopamine can regulate the intensity of phasic dopamine by the effect of the former on extracellular dopamine levels (Grace, 1991). On the whole, extensive literature exists on the question of specific roles of phasic and tonic DA, a comprehensive theory of the role of these two forms of release on various cortical and subcortical targets continues to be elusive. In the present paper we do not distinguish between these two forms of DA release. However, the single $\delta$ variable shown in this paper is similar to TD error, and therefore closer to phasic DA than tonic DA.

In simulations of normal reaching of Section 4.1, there is an early stage when arm exhibits prolonged, wandering movements before it reaches the target. When the arm approaches the target accidentally, a reward is delivered, which is used to train the MC. Reaching movements become more direct and brief as training proceeds. Thus variability in reaching trajectories decreases with learning (Fig. 6). Studies with primate reaching patterns reveal an exponential reduction of variability with learning (Georgopoulos et al, 1981). The exploratory movements of the arm, driven by the chaotic activity of indirect pathway, are reminiscent of the notion of motor babbling proposed in the context of imitation learning in infants (Meltzoff and Moore, 1997). Infants are thought to make random movements, and, by confirmatory feedback from an adult of the environment, learn to relate the movements initiated and the end states of the body. A similar learning of articulatory-auditory relation also seems to be driven by the more familiar vocal babbling (Kuhl and Meltzoff, 1996).

In the PD reaching simulations of Section 4.2, we considered three types of models. Typically PD pathophysiology is modeled purely in terms of reduction in dopamine. However, in the Type A PD model, we incorporated two factors related to PD pathology: 1) dopamine reduction, and 2) reduction in complexity of indirect pathway dynamics. Measures of impaired reaching movement like a) (longer) reaching duration indicating bradykinesia, 2) tremor and 3) undershoot are calculated. All the three types of PD models (A, B and C) showed longer reaching duration, more tremor, and greater undershoot compared to normal. In Type A, as disease progressed (increased loss of DA cells ($P_{DA}$), and decreased complexity of indirect pathway dynamics), these measures gradually approached an extreme value before they saturated. In Type B model, the measures showed a step-like variation. Type C model exhibits a smooth variation of symptoms, except for tremor, which exhibits a nearly flat peak.

The pattern of variation of symptoms seems to be reflected in the variation of times spent in various regimes. With increasing DA loss, the time spent in Go regime reduces, and δ spends more time within the Explore range. Thus time spent in Explore regime increases. This increase continues until δ falls below a critical value and enters the NoGo regime and remains mostly confined to the same. Thus we see that, as DA loss increases, the time spent in Explore regime gradually approaches a peak, and falls to near zero subsequently (in Types A and B). Since there is no restriction on δ in Type C, the time spent in Explore regime increases monotonically. In general, variation of time spent in the Explore regime seems to resemble variation of tremor, while undershoot and average velocity variation seem to follow the variation of time spent in Go regime. Experimental studies have linked tremor to changes in GP (Hurtado et al, 1999) and STN (Hamani et al 2004), in other words to changes in indirect pathway. In cases of akinetic-rigidity, Albin et al (1990) found a profound a loss of striatal cells projecting to GPi, which constitute the direct pathway. Based on features of disease progression in Huntington's disease, another neurodegenerative disorder, it was suggested that degeneration of direct pathway is responsible for rigidity and bradykinesia (Berardelli et al, 1999). In our model, in all the three types, the frozen, or rigid-like state is associated with drastically reduced times spent in Go regime, whose substrate is the direct pathway.

Increased movement duration in PD patients is a well-known clinical fact. In a study in which patients were asked to look and point to visual targets on a screen, PD patients took 24% more time to execute the movement than control subjects (Desmurget et al, 2003). Bradykinesia is thought to occur due to failure of BG output to reinforce the cortical mechanisms that prepare and execute the commands to move (Berardelli et al, 2001). In our model too, bradykinesia is a result of impaired interaction between BG and MC, which in turn is caused by DA cell loss.

Undershooting target is another prominent feature of goal-oriented PD movements. In a study in which PD patients were asked to copy target lines of fixed size, compared to controls, patients undershot the required size when the target size is greater than or equal to 2 cm (van Gemmert et al, 2003). It is noteworthy that, in PD patients, saccadic movements also typically undershoot targets, particularly in the vertical direction (White et al, 1983). In the simulations of the previous section, error in reaching includes both undershoot and error in direction. However, reaching error in PD patients is typically dominated by undershoot with no significant error in direction. This discrepancy in model performance related to differential error in reaching direction and undershoot will have to be investigated further in the future.

Tremor is another classic symptom of PD motor impairment. Our model too exhibits "tremor" which, however, might be different from the way it is characterized in experimental literature. Tremor in movement disorder literature is marked by presence of strong oscillatory components in electromyogram (EMG) and PD tremor is sometimes found to correlate with abnormal neural activity in GP (Hurtado et al, 1999) and STN (Hamani et al, 2004). In our model, "tremor" is quantified as root-mean-square (RMS) value of acceleration of the arm's end-effector (Appendix-I). Thus what we refer to as tremor, strictly speaking, denotes fluctuations in movement velocity, which is higher in the PD version of the model than in normal condition. Furthermore, the tremor described in our model emerges during reaching and is therefore akin to action tremor. Although action tremor is found in PD patients, resting tremor is found more often than action tremor. One way of extending the current model to address the problem of resting tremor is to treat the "resting state" as another possible target location. Since a typical hand may be assumed to spend more time in the resting state that in any other state, this feature can be incorporated in the simulations. It would be interesting to note the differences in the "action tremor" and "resting tremor" that emerge from such a model.

Another aspect of tremor in the model is that in the present work, we use a simple measure of tremor based on RMS value of acceleration, but, considering the suggested link with degradation of chaos in the indirect pathway, it would perhaps be more appropriate to perform chaotic time-series analysis on tremor and check if there is a reduction in chaoticity with disease progression. Such analyses will have a bearing on clinical data since it was reported that PD tremor displays reduced chaoticity due the effect of treatment (Yulmetyev et al, 2006). These alternative measures of tremor will form part of future work.

Dounskaia et al (2009) characterize movement irregularities in PD reaching using a measure called Normalized Jerk Score (NJS) and show that NJS in PD patients is greater than in normal subjects. Since PD movements are known to have abnormal fluctuation in velocity and acceleration, the magnitude of jerk, which refers to temporal derivative of acceleration, is understandably higher in PD patients than in normals. The proposed model is a lumped model of BG, which aims to embody the essence of BG dynamics. It is meant to present a picture of BG in which the direct pathway subserves exploitation and indirect pathway subserves exploration (Fig. 30) and in this respect departs radically from existing BG modeling literature. It only attempts to present the large-scale picture, and is not meant to be a detailed, network-level or biophysical model of BG function. It is a systems-level model, which aims to link PD pathology at circuit level with its behavioral manifestations in reaching. To achieve such a wide scope, model components have been simplified. The arm and the muscles involved are static models, and therefore arm dynamics are produced purely by temporal variation in muscle activation. A more realistic model would use a dynamic arm, and also incorporate a forward model necessary to control the arm. The actor/MC is also a static model, a perceptron, which happens to be adequate for the problem at hand. The Value function is pre-calculated and is not trained by δ, as it should be in a full RL framework. There is also no explicit representation of striatum nor are there corticostriatal connections modifiable by DA signals. A novel feature of the present model is to represent part of the indirect pathway dynamics using a chaotic system, and suppress its chaoticity to represent PD pathology. We envision two stages of future development of this model. In the first, each of the model components is replaced by networks of abstract neurons with appropriate dynamics. The second stage would consist of biophysical neuron models, with the model architecture closely complying BG anatomy.

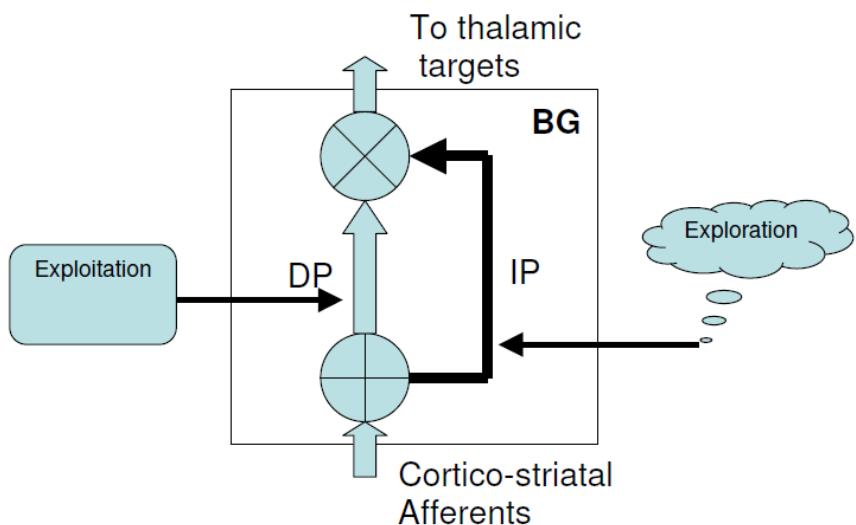

Figure 30: A hypothetical schema of BG function in which direct pathway subserves exploitation and indirect pathway exploration

Another novel feature of the proposed model is that it attempts to capture disease progression in a neurodegenerative disorder (NDD) like PD, as opposed to contrasting normal function with disease state at a particular level. NDDs, whose incidence seems to be increasing dramatically, are marked by a progressive impairment in function. Understanding the nature of this progressive impairment is further complicated by the fact that the impairment is usually associated with high short-term fluctuation in symptoms (Walker et al, 2000). Although neurological deficits in such cases are thought to be related to neuronal cell loss, more recent findings suggest that the situation could be more complicated (Terry et al, 1991). Behavioral impairment in NDDs is associated with formation of abnormal protein assemblies (plaques, tangles and inclusion bodies), neuronal cell loss and also network dysfunction (Palop et al, 2006). An integral understanding of disease progression entails progress in understanding at all the above levels. Another tricky affair in NDDs is to be able to distinguish between a co-pathologic and a compensatory change. Only an integral understanding of NDD progression at network level helps development of effective therapies for NDD. The present model marks a step in that direction for the specific case of PD.

**STN-GPe and Exploration:**

A key idea that is embodied in the present model, - an idea that is developed from earlier work (Sridharan et al, 2006; Gangadhar et al, 2008) – is the idea that the STN-GPe system, which constitutes the indirect pathway, plays the role of the Explorer in BG dynamics. RL-based or Actor-Critic models are an important class of models describing BG function. Of the three key components of RL, viz., Actor, Critic and Explorer, substrates to both Actor and Critic have been located within BG nuclei; however, no subcortical substrate to the Explorer has been discovered in experimental work, nor suggested in modeling studies. Functional imaging studies identify two cortical substrates of exploration - Anterior frontopolar cortex and Intraparietal Sulcus – but no subcortical counterpart of exploration has been found (Daw et al, 2006). On the other hand the roles attributed to the STN-GPe have been variable and tentative ranging from movement inhibition (Albin et al, 1989; Frank, 2006), focusing and sequencing (Hikosaka et al, 2000), action selection (Gurney, 2001), switching (Isoda and Hikosaka, 2008). Thus we try to fit the peg of a missing subcortical substrate for exploration into the hole of a tentative understanding of STN-GPe function, and propose that the STN-GPe system *is* the subcortical substrate for exploration.

The STN-GPe loop is often studied as a single unit perhaps since oscillations produced by this loop have fascinated many researchers (Terman et al, 2002). Based on their studies of BG organotypic tissue cultures, Plenz and Kitai (1999) have proposed that correlated activity can arise in both STN and GPe structures and is caused by the interaction between the two structures rather than being driven by an external source. Recent experimental studies have revealed prominent low-frequency periodicity (4-30Hz) of firing and dramatically increased correlations among neurons in the GPe and the STN, though there were no significant changes in firing rates (Bergman et al, 1994; Nini et al, 1995; Magnin et al, 2000; Brown et al, 2001). Under dopamine-deficient conditions associated with PD, recordings from STN neurons of PD animals and patients revealed synchronized oscillations (Magnin et al., 2000; Nini et al, 1995).

Thus we propose a functional role to the presence of complex oscillations in STN-GPe in normal conditions, and explain the pathological consequences of the loss of complex dynamics in that structure. The idea of explaining PD symptoms in terms of reduction in complexity of dynamics of relevant neural structures had existed for some time (Edwards, Beuter and Glass, 1999). It is from such considerations that PD had been dubbed a "dynamical disease" (Beuter and Vasilakos, 1995). Accordingly, fixed point dynamics had been linked to akinetic rigidity of PD, and limit cycles to PD tremor. Several instances have been discovered in physiology, particularly cardiac physiology, where chaotic activity of a system is essential for its normal function (Goldberger, Rigney, & West, 1990). We may have a similar situation in the STN-GPe system: complex activity may correspond to normal function and loss of complexity to disease.

Walker, M.P., Ayre G.A., Cummings J.L., et al. (2000). Quantifying fluctuation in dementia with Lewy bodies, Alzheimer's disease, and vascular dementia. *Neurology*, Vol. 54, pp1616-1625.

White, O.B., Saint-Cyr J.A., Tomlinson, R.D., Sharpe J. (1983). Ocular motor deficits in Parkinson's disease. II: Control of saccadic and smooth pursuit systems. *Brain*, Vol. 106, pp 571-587.

**Appendix I:**

<u>Normal:</u>

1) **Average Velocity**:

> It is computed as the rate of change of displacement at the end of a reach.

2) **Actor Error**:

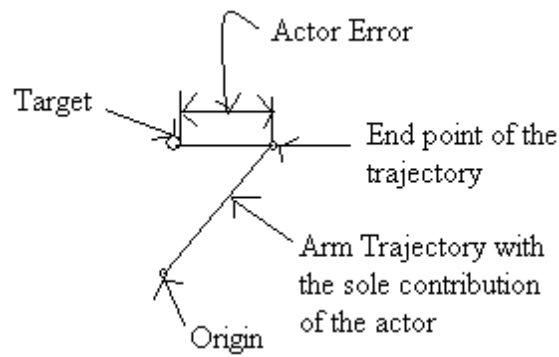

> It is defined as the magnitude of the vector connecting the target and the end point reached by the arm with the sole contribution of the actor.

3) **Path Variability**:

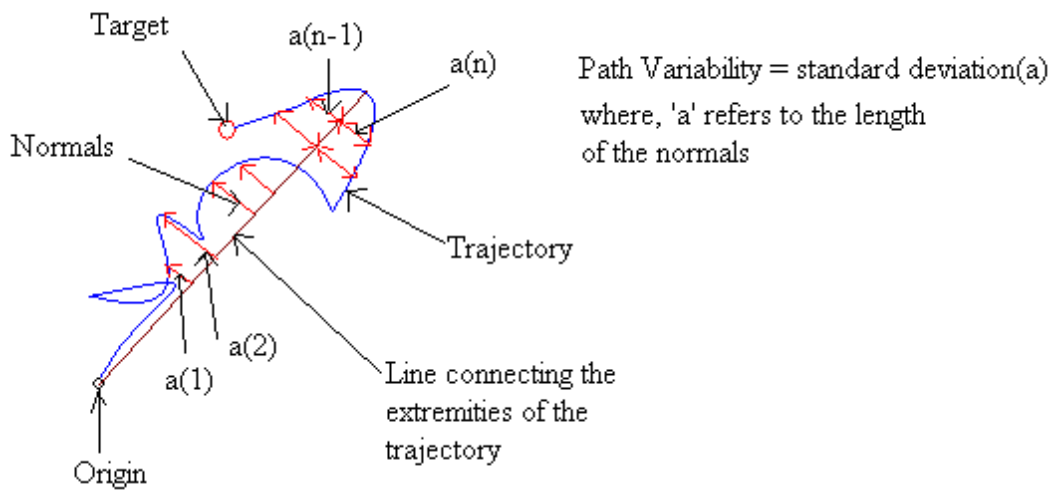

Path Variability = standard deviation(a)
where, 'a' refers to the length
of the normals

Path Variability of the reach trajectory is defined as the standard deviation of the length of the normals from the line connecting its extremities, intersecting it.

PD:

1) **Undershoot factor:**

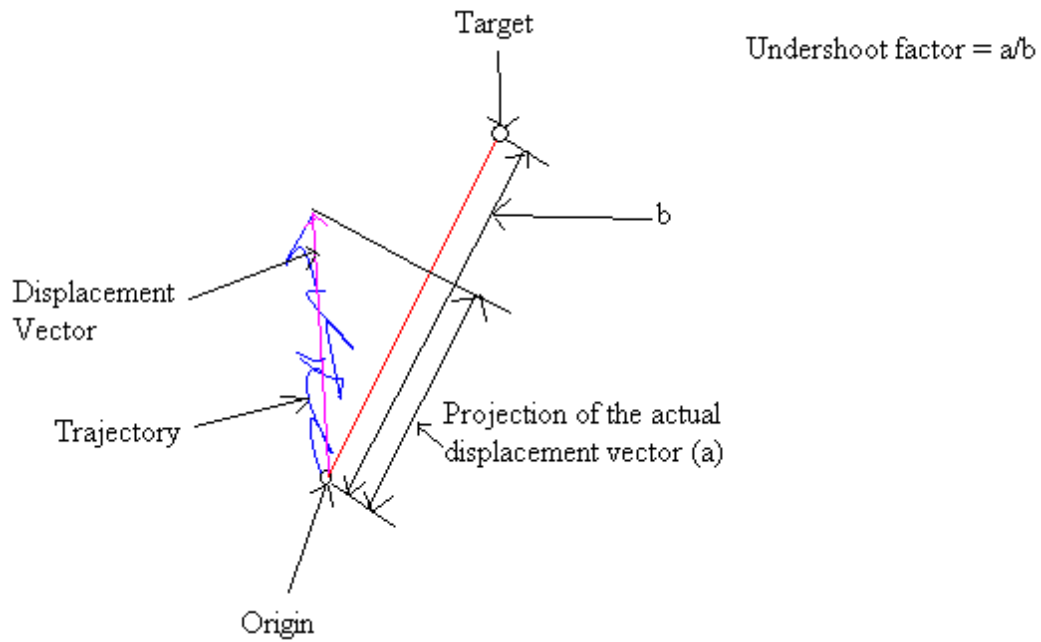

Undershoot factor = a/b

It is defined as the ratio of the magnitudes of the projection of the actual displacement vector of the reach onto the vector connecting the origin to the target, to that of the vector connecting the origin to the target.

2) **Tremor Factor:**

It is defined as the root mean square value of the acceleration of the arm during the reaching task.

3) **Average velocity**:

Same as the normal case

**Appendix II**: An altered form of dynamics in *NoGo* regime.

Of the three regimes described by eqns. (3.5), in the NoGo regime, the BG output at time 't' is changed such that it is negative of the change that occurred in the previous step (t-1). Since the traditional definition of NoGo is to withhold movement and not reverse movement, we consider an alternative form of NoGo dynamics in which there is no change in BG output. Thus the new regimes can be defined as follows:

$$if\,(\delta(t) > DA_{hi})$$
$$\Delta g_{bg}(t) = +\Delta g_{bg}(t-1) \qquad -"Go"$$
$$else\,if\,(\delta(t) > DA_{lo} \quad and \quad \delta(t) \le DA_{hi})$$
$$\Delta g_{bg}(t) = \varphi \qquad -"E\,\mathrm{xp}\,lore" \qquad\qquad\qquad (A2.1)$$
$$else\,/\,/\;\,(\delta(t) \le DA_{lo})$$
$$\Delta g_{bg}(t) = 0 \qquad -"NoGo"$$

where φ is defined as before in eqns. (3.5). With this variation of the NoGo regime, we repeat the simulations of the normal case described in Section 4.1. The results obtained are as follows.

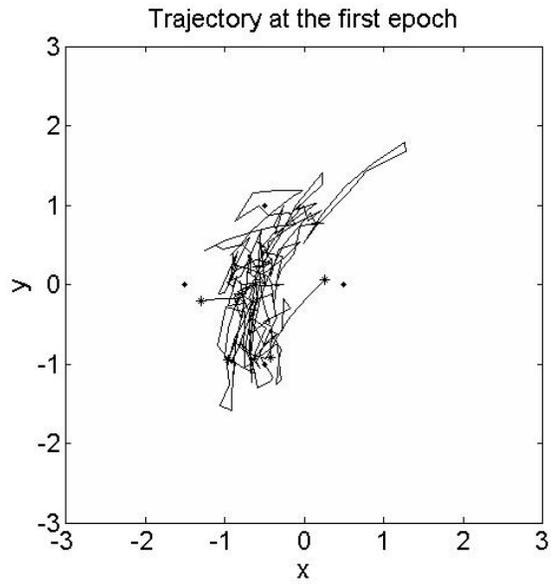

Figure A2.1: Trajectory during the first epoch (Under altered NoGo form of dynamics)

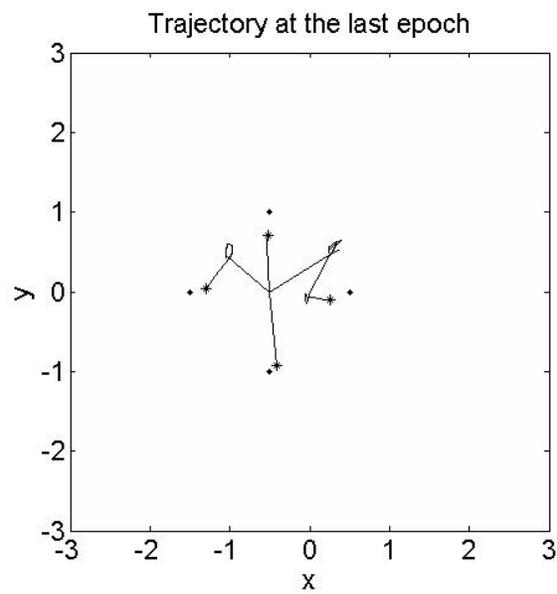

Figure A2.2: Trajectory at the last epoch (Under altered NoGo form of dynamics)

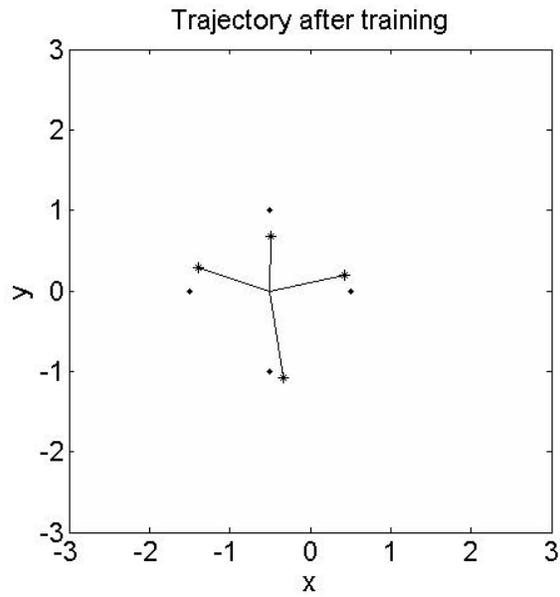

Figure A2.3: Trajectory after training (MC alone, no BG contribution, Under altered NoGo form of dynamics)

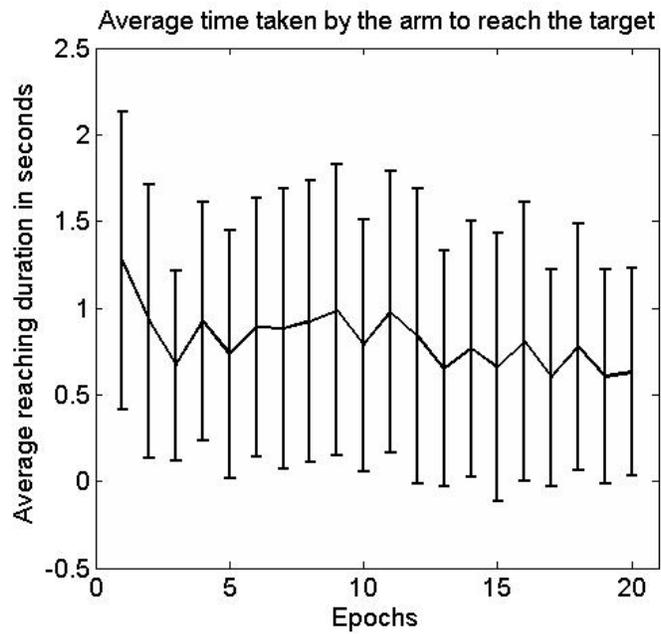

Figure A2.4: Variation of Average Reaching Duration with learning in normal conditions (Under altered NoGo form of dynamics)

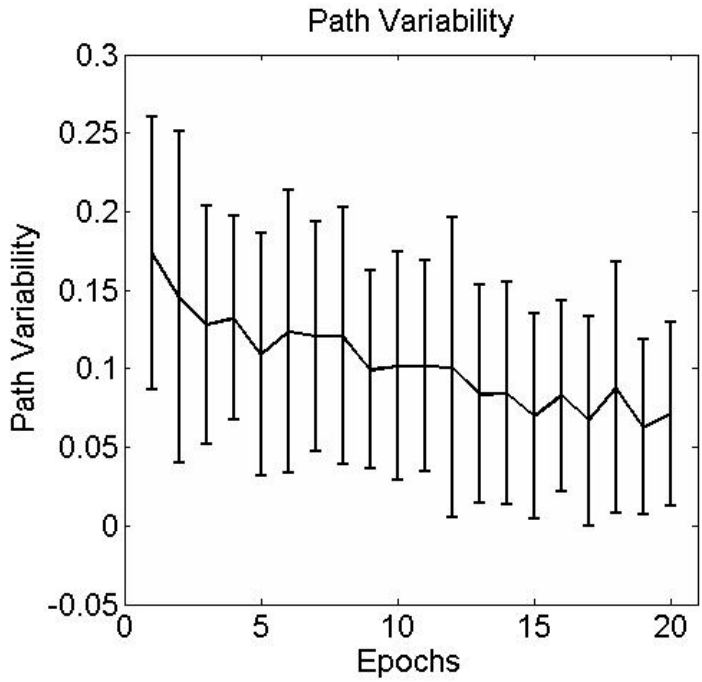

Figure A2.5: Variation of Path Variability with learning in normal conditions. (Under altered NoGo form of dynamics.)

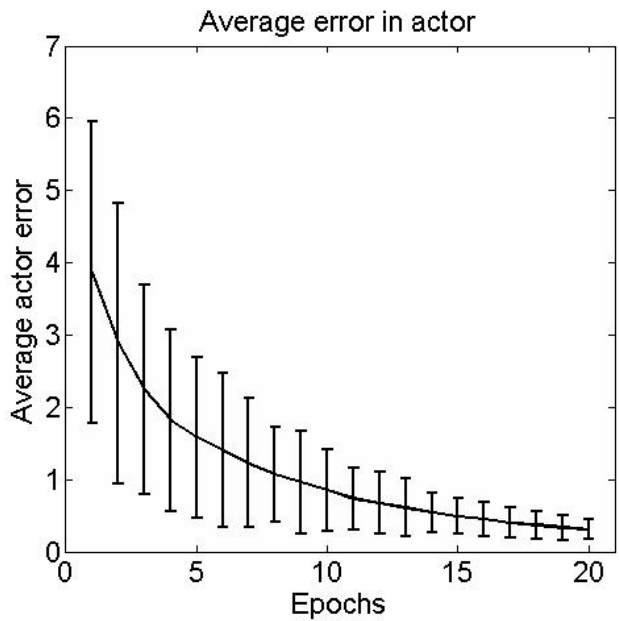

Figure A2.6: Variation of output error of MC with learning in normal conditions. (Under altered NoGo form of dynamics)

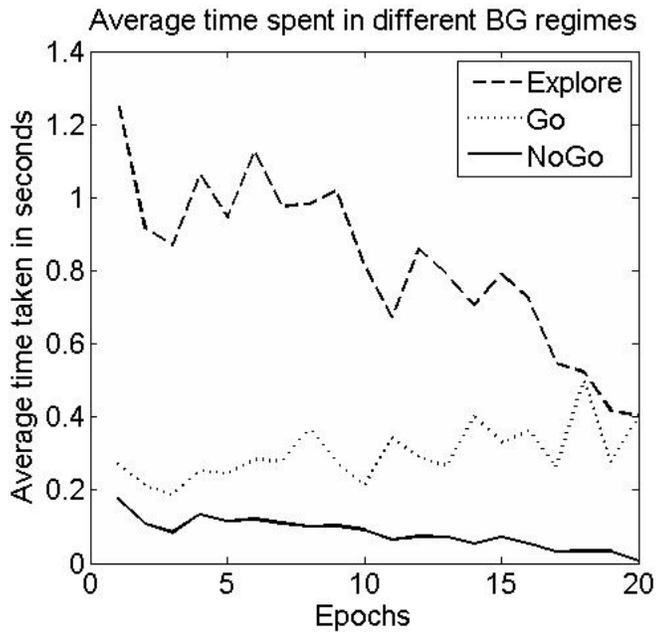

Figure A2.7: Average time spent in various BG regimes (Under altered NoGo form of dynamics)

Comparing Figs. A2.1-7 with Figs. 2 - 8 of Section 4.1, we note that most results are similar except one significant difference. The reaching time depicted in Fig. A2.4 is somewhat longer than the reaching times depicted in Fig. 5. Therefore, for the pathological studies, we use only the NoGo regime as it is depicted in eqns. (3.5) and do not pursue the NoGo regime as it is depicted in eqns. A2.1.